\documentclass[aps,twocolumn,showpacs]{revtex4}

\usepackage{graphicx}
\usepackage{amsfonts}
\usepackage{amssymb}
\usepackage{amsbsy}
\usepackage{amsmath}
\usepackage{latexsym}
\usepackage{natbib}
\usepackage{bm}
\usepackage{subfigure}
\usepackage{color}
\usepackage{lscape}

\renewcommand*{\l}{\lambda_{\star}}

\def\ce{\mathrm{ce}}
\def\se{\mathrm{se}}

\def\k{\mathrm{k}}
\def\kr{\kappa}

\def\x{x}
\def\y{y}
\def\xir{\xi}

\def\Vr{\bar{V}}
\def\Hr{\bar{H}}

\def\omegar{\bar{\omega}}

\def\H{\mathcal{H}}

\def\vacl{\Theta}

\begin{document}

\title{Absence of Unruh effect in polymer quantization}

\author{Golam Mortuza Hossain}
\email{ghossain@iiserkol.ac.in}

\author{Gopal Sardar}
\email{gopal1109@iiserkol.ac.in}

\affiliation{ Department of Physical Sciences, 
Indian Institute of Science Education and Research Kolkata,
Mohanpur - 741 246, WB, India }
 

\date{\today}

\begin{abstract}

Unruh effect is a landmark prediction of standard quantum field theory in which
Fock vacuum state appears as a thermal state with respect to an uniformly
accelerating observer.  Given its dependence on trans-Planckian modes, Unruh
effect is often considered as an arena for exploring a candidate theory of
quantum gravity. Here we show that Unruh effect disappears if, instead 
of using Fock quantization, one uses polymer quantization or loop quantization, 
the quantization method used in loop quantum gravity. Secondly, the polymer 
vacuum state remains a vacuum state even for the accelerating observer in the 
sense that expectation value of number density operator in it remains zero. 
Finally, if experimental measurement of Unruh effect is ever possible then it 
may be used either to verify or rule out a theory of quantum gravity.

\end{abstract}

\maketitle

\section{Introduction}

A challenging problem for any theory of quantum gravity is to make physical
predictions which are within the reach of current experiments either directly or
indirectly. This problem originates from the fact that the energy scale
accessible even by modern experiments are too small compared to the scale of
quantum gravity namely the \emph{Planck scale}. On the other hand, for
trans-Planckian modes, such a theory often makes significantly different
physical predictions, as compared to those from standard \emph{quantum field
theory} and \emph{general relativity}.  For low-energy modes, such Planck-scale
corrections are usually very small. Therefore, in order to confront a theory
that modifies mainly Planck scale physics, one is often forced to look for
physical phenomena which are dependent on the trans-Planckian modes yet may have
physical implications in a relatively low energy regime.

Unruh effect \cite{Fulling:1972md, Unruh:1976db,Crispino:2007eb} is 
an intriguing consequence of standard quantum field theory in a
curved background \cite{Birrell1984quantum}, in which an \emph{uniformly
accelerating} observer finds Fock vacuum state to be a \emph{thermal state}.
In order to obtain this thermal
spectrum, as seen by the accelerating observer, one needs to include
contributions even from the trans-Planckian modes, as seen by an \emph{inertial
observer}. This particular feature makes Unruh effect to be a potentially
important arena for understanding and exploring the implications of Planck-scale
physics \cite{Padmanabhan:2009vy}.

Polymer quantization \cite{Ashtekar:2002sn,Halvorson-2004-35}, the quantization
method used in loop quantum gravity
\cite{Ashtekar:2004eh,Rovelli2004quantum,Thiemann2007modern}, differs from
Schrodinger quantization in several important ways when applied to a mechanical
system. Firstly, apart from \emph{Planck constant} $\hbar$, it contains a new
dimension-full parameter. In the context of quantum gravity, this new scale
would correspond to \emph{Planck length} $L_p = \sqrt{\hbar G/c^3}$,  where $G$
is Newton's constant of gravitation and $c$ is the speed of light. Secondly,
instead of both, only one of the position and momentum is represented directly
as an \emph{elementary operator} in the kinematical Hilbert space. The second
operator is represented as exponential of its classical counterpart. These
features together with a \emph{distinct} kinematical inner product make polymer
quantization unitarily \emph{inequivalent} to Schrodinger quantization
\cite{Ashtekar:2002sn}. Therefore, in principle, polymer quantization can lead
to a different set of results compared to those from Schrodinger quantization.

In the standard derivation of Unruh effect \emph{i.e.} using Fock quantization,
the field operator is expressed in terms of \emph{creation} and
\emph{annihilation} operators of different Fourier modes. Each of these modes
behaves as a \emph{mechanical} system corresponding to a decoupled harmonic
oscillator and is quantized using Schrodinger method. In polymer quantization of
these modes, the notions of creation and annihilation operators are \emph{not}
available. Therefore, we employ here a new method in which Unruh effect is
derived using \emph{energy spectrum} of these modes for both Fock and polymer
quantizations. Subsequently, we show that Unruh effect disappears in 
the case of polymer quantization when one removes the associated integral 
regulator. We also comment on certain criticism and discuss few key issues 
related to the above result.

\section{Rindler spacetime}

The spacetime as seen by an \emph{uniformly accelerating} observer using
\emph{conformal} Rindler coordinates $\bar{x}^{\alpha} = (\tau,\xi,y,z)$ in
\emph{natural} units ($c=\hbar=1$), is described by the metric
\cite{Rindler:1966zz}
\begin{equation}
 \label{RindlerMetric}
  ds^2 = e^{2a\xi} \left( -d\tau^2 + d\xi^2 \right) + dy^2 + dz^2
  \equiv g_{\alpha\beta}d\bar{x}^{\alpha} d\bar{x}^{\beta} ~,
\end{equation}
where parameter $a$ denotes the magnitude of \emph{acceleration} 4-vector. For
comparison, we consider an \emph{inertial} observer who uses Minkowski
coordinate system $x^{\mu} = (t,x,y,z)$ together with the metric $ds^2 =
\eta_{\mu\nu}dx^{\mu}dx^{\nu} = - dt^2 + dx^2 + dy^2 + dz^2$.
If the accelerating observer \emph{i.e.} Rindler observer, moves along $+ve$
x-axis with respect to the inertial observer then their respective coordinates
are related to each other as
\begin{equation}
 \label{RindlerMinkowskiRelation}
 t =  \frac{1}{a} e^{a\xi} \sinh a\tau ~,~~~
 x =  \frac{1}{a} e^{a\xi} \cosh a\tau ~.
\end{equation}
One may note that Rindler spacetime covers only a wedge-shaped region of
Minkowski spacetime. This region is known as \emph{Rindler wedge}. The $y$ and
$z$ coordinates between these two observers are trivially related. So for
simplicity we restrict ourselves to an $(1+1)$ dimensional spacetime and denote
their respective coordinates as $(\tau,\xir)$ and $(t,\x)$ for further study.

\section{Massless scalar field}

The dynamics of a massless scalar field in $(1+1)$ dimensional Minkowski
spacetime is described by the action
\begin{equation}
 \label{ScalarActionMinkowski}
 S_{\Phi} = \int dtd\x \left[ - \frac{1}{2} \sqrt{-\eta} \eta^{\mu\nu}  
 \partial_{\mu} \Phi(t,\x)  \partial_{\nu} \Phi(t,\x) \right] ~.
\end{equation}
For Rindler observer the scalar field action is 
$S_{\Phi} = \int d\tau d\xir [- \frac{1}{2} \sqrt{-g} g^{\alpha\beta}  
\partial_{\alpha} \Phi(\tau,\xir)\partial_{\beta} \Phi(\tau,\xir)]$.
Given the \emph{conformal} structure of Rindler metric $g_{\alpha\beta}$, the
action can also be expressed as 
$S_{\Phi} = \int d\tau d\xir [ - \frac{1}{2} \sqrt{-g^{0}} {g^0}^{\alpha\beta}  
  \partial_{\alpha} \Phi(\tau,\xir) \partial_{\beta} \Phi(\tau,\xir) ]$
where $ g_{\alpha\beta}(\tau,\xi) = e^{2a\xi} g^0_{\alpha\beta}(\tau,\xi)$. 
In other words, for Rindler observer with the coordinates $(\tau,\xi)$, 
dynamics of a massless scalar field can be described \emph{equivalently}
using the metric $g^0_{\alpha\beta} = diag(-1,1)$. It allows one to perform 
Fourier transformation similar to that of an inertial observer.
The Hamiltonian corresponding to the action (\ref{ScalarActionMinkowski}) is
given by
\begin{equation}\label{SFHamGen}
H_{\Phi}  =  \int d\x \left[ \frac{\Pi^2}{2\sqrt{q}} +
\frac{\sqrt{q}}{2} q^{ab} \partial_a\Phi \partial_b\Phi
\right] ~,
\end{equation}
where $q_{ab}$ is metric on the \emph{spatial} hyper-surfaces. The Poisson
bracket between the field $\Phi$ and the conjugate momentum $\Pi$ is 
\begin{equation}\label{PositionSpacePB}
\{\Phi(t,\x), \Pi(t,\y)\} = \delta(\x-\y) ~.
\end{equation}
For Rindler observer, one can write down a similar scalar field Hamiltonian.

\subsection{Fourier modes}

We define Fourier modes for the scalar field and its momentum with respect to
the inertial observer as
\begin{equation}\label{Minkowski:FourierModesDef}
\Phi = \frac{1}{\sqrt{V}} \sum_{\k} \tilde{\phi}_{\k}(t) e^{i {\k} {\x}} ,~
\Pi  = \frac{1}{\sqrt{V}} \sum_{\k} \sqrt{q} ~\tilde{\pi}_{\k}(t) 
e^{i {\k} {\x}},
\end{equation}
where $V=\int d\x \sqrt{q}$ is the spatial volume. For Minkowski spacetime, the
spatial volume would diverge as the space is non-compact. In order to avoid
dealing with divergent quantity, we consider a fiducial box of finite volume. In
particular, the volume of the box is explicitly chosen as
\begin{equation}
\label{MinkowskiVolume}
V = \int_{L^{-}}^{L^{+}} d\x \sqrt{q} = L^{+} - L^{-} \equiv L ~.
\end{equation}
For brevity of notation, we shall skip explicitly writing the limits of the
integration.
In this context, Kronecker delta and Dirac delta can be expressed as
$\int d\x \sqrt{q} ~e^{i (\k-\k') \x} = V \delta_{\k,\k'}$ and 
$\sum_{\k} e^{i \k (\x-\y)} = V \delta (\x-\y)/\sqrt{q}$. These expressions
together imply that $\k \in \{\k_r\}$ where $\k_r = (2\pi r/L)$ with $r$ being
any \emph{non-zero integer}.

In terms of the Fourier modes (\ref{Minkowski:FourierModesDef}), the Hamiltonian
(\ref{SFHamGen}) can be reduced to $H_{\Phi} = \sum_{\k} \H_{\k}$ where
Hamiltonian density is
\begin{equation}\label{SFHamFourierMinkowski}
\H_{\k} = \frac{1}{2} \tilde{\pi}_{-\k} \tilde{\pi}_{\k} +
\frac{1}{2} |\k|^2 \tilde{\phi}_{-\k}\tilde{\phi}_{\k}  ~,
\end{equation}
and momentum-space Poisson bracket is
\begin{equation}\label{Minkowski:MomentumSpacePB}
\{\tilde{\phi}_{\k}, \tilde{\pi}_{-\k'}\} =  \delta_{\k,\k'}
~.
\end{equation}
Being a \emph{complex-valued} function, each $\tilde{\phi}_{\k}$ has two
independent modes.  However, the scalar field $\Phi$ being a \emph{real-valued}
function not all $\tilde{\phi}_{\k}$'s are independent. In particular, reality
of the scalar field requires $\tilde{\phi}_{\k}^{*}  = \tilde{\phi}_{-\k}$. We
shall impose this reality condition before quantizing these modes.

We define Fourier modes for Rindler observer in a similar manner as
$\Phi(\tau,\xir) = (1/\sqrt{\bar{V}})\sum_{\kr} \tilde{\phi}_{\kr}(\tau) e^{i
{\kr} {\xir}}$ and $\bar{\Pi}(\tau,\xir)  = (1/\sqrt{\bar{V}}) \sum_{\kr}
\sqrt{\bar{q}}~\tilde{\pi}_{\kr}(\tau) e^{i {\kr} {\xir}}$ where the spatial
volume $\Vr = \int d\xir \sqrt{\bar{q}}$.  Here $\bar{q}$ is the determinant of
the spatial metric corresponding to the flat spacetime metric
$g^0_{\alpha\beta}$.
As earlier, the scalar field Hamiltonian for Rindler observer can be reduced to
$\Hr_{\Phi} = \sum_{\kr}\bar{\H}_{\kr}$ where 
\begin{equation}\label{SFHamFourierRindler}
\bar{\H}_{\kr} = \frac{1}{2} \tilde{\pi}_{-\kr} \tilde{\pi}_{\kr} +
\frac{1}{2} |\kr|^2 \tilde{\phi}_{-\kr} \tilde{\phi}_{\kr} ~,~
\{\tilde{\phi}_{\kr}, \tilde{\pi}_{-\kr'}\} =  \delta_{\kr,\kr'} ~.
\end{equation}

\subsection{Relation between Fourier modes of two observers}

The scalar field is invariant under coordinate transformation \emph{i.e.}
$\Phi(\tau,\xir)=\Phi\left(t(\tau,\xi), \x(\tau,\xi)\right)$.  On the other
hand, respective canonical momenta $\Pi(t,\x) = \partial \Phi(t,\x)/\partial t$
and $\bar{\Pi}(\tau,\xi) = \partial \Phi(\tau,\xi)/\partial \tau$ can be related
to each other using
$\partial \Phi(\tau,\xi)/{\partial \tau} = 
({\partial t}/{\partial\tau})({\partial \Phi(t,\x)}/{\partial t}) + 
({\partial \x}/{\partial\tau})({\partial \Phi(t,\x)}/{\partial \x})$.
We may recall that in canonical formulation, the spacetime is broken into
spatial hyper-surfaces and these hyper-surfaces are labeled by different
instances of time. Given an initial field and momentum configuration on a
particular hyper-surface, it is possible to dynamically evolve to any other
hyper-surface uniquely. Therefore, for simplicity but without loss of
generality, we choose the spatial hyper-surface labeled by $t=\tau=0$ for making
comparison between these two observers. At $\tau=0$, $\partial x/\partial\tau=0$
and the spatial coordinates $x$ and $\xi$ are related to each other as $a x =
e^{a\xi}$. The spatial volume $\Vr$ can be expressed as $a \Vr = \ln
\left(L^{+}/L^{-}\right)$. The Fourier modes for Rindler observer can be
expressed in terms of the modes of the inertial observer as
\begin{eqnarray}
\label{FourierFieldRelation}
\tilde{\phi}_{\kr} &=&  \sum_{\k >0} \tilde{\phi}_\k F_0(\k,- \kr) + 
\sum_{\k >0} \tilde{\phi}_{-\k} F_0(-\k,- \kr) ~, \nonumber \\
\tilde{\pi}_{\kr} &=& \sum_{\k >0} \tilde{\pi}_\k F_1(\k,-\kr) + 
\sum_{\k>0} \tilde{\pi}_{-\k} F_1(-\k,-\kr) ~,
\end{eqnarray}
where $\tilde{\phi}_{\kr} = \tilde{\phi}_{\kr}(0)$, $\tilde{\phi}_\k =
\tilde{\phi}_\k(0)$, $\tilde{\pi}_{\kr} = \tilde{\pi}_{\kr}(0)$,
$\tilde{\pi}_{\k}= \tilde{\pi}_{\k}(0)$. Given $\sqrt{\bar{q}}=1$, the
coefficients $F_0$, $F_1$ can be written as
\begin{equation}\label{FmDef}
 F_m(\k,\kr) =  \frac{1}{\sqrt{V \Vr}} \int d\xi ~e^{m a \xi} ~e^{i \k \x + i 
\kr \xi} ~,
\end{equation}
for $m=0,1$. These coefficient functions are analogous to standard
\emph{Bogoliubov coefficients}.

\subsection{Regularization of the coefficient functions}

The definition of $F_m(\k,\kr)$ implies that $F_1(\k,\kr) = (-i a)\partial
F_0(\k,\kr)/\partial \k$. Clearly, knowing the expression of $F_0(\k,\kr)$ is
sufficient. However, the integrand being a pure phase, $F_0(\k,\kr)$ does not
\emph{converge} when volume regulators are removed. In order to avoid dealing
with formally divergent terms, we introduce a non-oscillatory \emph{regulator}
term in the expression of $F_m(\k,\kr)$ as follows
\begin{equation}\label{FmDeltaDef}
 F_m^{\delta}(\k,\kr) =  \frac{1}{\sqrt{V \Vr}} \int d\xi ~e^{m a \xi} 
 ~e^{i \k \x + i \kr \xi} \left[ \frac{e^{\delta a \xi}}{d_m} \right] ~,
\end{equation}
where $d_m = (1- i\delta m a/\kr)$. In the limit $\delta \to 0$,
$F_m^{\delta}(\k,\kr)$ reduces to $F_m(\k,\kr)$.
The change of variable $u \equiv |\k|x$ in the integration (\ref{FmDeltaDef})
would lead to
\begin{equation}\label{FmDeltaE1}
 F_m^{\delta}\left(\pm|\k|,\kr\right) =  
  \frac{a^{\beta} |\k|^{-\beta-1}}{\sqrt{V \Vr}~d_m} 
  I_{\pm}\left(\beta\right) ~,
\end{equation}
where $\beta = (i\kr/a + \delta + m - 1)$ and $I_{\pm}(\beta) = \int du ~e^{\pm
i u} u^{\beta}$.  Based on the \emph{sign} of $\k$, the integral can be
evaluated by analytic continuation in \emph{upper} or \emph{lower} half of the
complex plane respectively as
\begin{equation}\label{IpmBeta}
 I_{\pm}(\beta) = e^{\pm i\pi(\beta+1)/2} ~\Gamma(\beta+1) ~,
\end{equation}
where $\Gamma(\beta+1)$ denotes \emph{Gamma function}. To obtain a
\emph{closed-form} expression (\ref{IpmBeta}), we have added two extra terms to
it, namely $\Delta I^{-}_{\pm} \equiv \int_0^{|\k|L^{-}} du ~e^{\pm i u} 
u^{\beta}$ and $\Delta I^{+}_{\pm} \equiv \int_{|\k|L^{+}}^{\infty}du ~e^{\pm i 
u} u^{\beta}$. Each of
these extra terms \emph{identically goes to zero} when volume regulators are
removed \emph{i.e.} when the limits $L^{-}\to0$ and $L^{+}\to\infty$ are
taken.
We note following two useful relations for different arguments of
$F_m^{\delta}(\k,\kr)$ as
\begin{eqnarray}
\label{F0:F0:Relation}
F_0^{\delta}\left(-|\k|,\kr\right) &=& e^{\pi\kr/a -i \delta\pi} 
~F_0^{\delta}\left(|\k|,\kr \right) ~,~\\
\label{F1:F0:Relation}
F_1^{\delta}\left(\pm|\k|,\kr\right) &=& \mp \frac{\kr}{|\k|}
F_0^{\delta}\left(\pm|\k|,\kr\right) ~.~
\end{eqnarray}
The requirement that both Poisson brackets $\{\tilde{\phi}_{\kr},
\tilde{\pi}_{-\kr}\}=1$ and $\{\tilde{\phi}_{\k}, \tilde{\pi}_{-\k}\} = 1$ are
simultaneously satisfied, demands $\sum_{\k>0} [ F_0(\k,-\kr)F_1(-\k,\kr) +
F_0(-\k,-\kr)F_1(\k,\kr) ] = 1$. Regulated expression of this condition which
remains \emph{real-valued} by the choice of $d_m$, requires
\begin{equation}
 \label{Volume-delta-restriction}
 \frac{(\kr/a) ~ |\Gamma(i\kr/a + \delta)|^2}
  {2\pi \left(e^{\pi\kr/a} - e^{-\pi\kr/a}\right)^{-1}}
 = \frac{(a\Vr) (2\pi/aV)^{2\delta}}{\zeta(1+2\delta)} ~,
 \end{equation}
where $\zeta(1+2\delta) \equiv \sum_{r=1}^{\infty} r^{-(1+2\delta)}$ is
\emph{Riemann zeta function}. In the limit $\delta \to 0$, \emph{lhs} of the
equation (\ref{Volume-delta-restriction}) becomes $1$ as $\Gamma(z)\Gamma(1-z) =
\pi /\sin \pi z$. Together with \emph{zeta function identity}
$\lim_{s\to0}[s~\zeta(1+s)] = 1$, the equation (\ref{Volume-delta-restriction})
requires $a L^{-} \simeq 2\pi ~ e^{-1/2\delta}$. In other words, consistency
of the regulated expression implies that the volume regulator $L^{-}$ and the
integral regulator $\delta$ are related and should be removed together as above.

\subsection{Hamiltonian densities of Fourier modes}

The Hamiltonian densities $\H_{\k}$ (\ref{SFHamFourierMinkowski}) and
$\bar{\H}_{\kr}$ (\ref{SFHamFourierRindler}) can be related to each other by
using the equations (\ref{FourierFieldRelation},\ref{Volume-delta-restriction})
for \emph{positive} $\kr$ \emph{i.e.} for $\kr>0$ as
\begin{equation} \label{Hamiltonian:Relation2}
\frac{\bar{\H}_{\kr}}{|\kr|} = \frac{h_{\kr}}{|\kr|} + 
\left(\frac{e^{2\pi\kr/a} + 1}{e^{2\pi\kr/a} - 1} \right)
\frac{1}{\zeta(1+2\delta)}
\sum_{r=1}^{\infty} \frac{\H_{\k_r}/|\k_r|}{r^{1+2\delta}},
\end{equation}
where 
$ h_{\kr} = \sum_{\k\neq\k'} [ \frac{1}{2} \tilde{\pi}_{\k} \tilde{\pi}_{-\k'}
  F_1^{\delta}(\k,-\kr) F_1^{\delta}(-\k',\kr) + \frac{1}{2}|\kr|^2
 \tilde{\phi}_{\k} \tilde{\phi}_{-\k'} 
 F_0^{\delta}(\k,-\kr) F_0^{\delta}(-\k',\kr)]$.
The expression of $h_{\kr}$ involves terms which are \emph{linear} in
$\tilde{\phi}_{\k}$'s and $\tilde{\pi}_{\k}$'s. Hence it will dropout from
vacuum expectation value.

\section{Number operator and vacuum state}

In our analysis, so far we have used only classical aspects of the
complex-valued mode functions $\tilde{\phi}_{\k}$ and $\tilde{\pi}_{\k}$. We now
redefine these modes in terms of real-valued functions $\phi_{\k}$ and
$\pi_{\k}$ such that the \emph{reality condition} of the scalar field $\Phi$ is
satisfied. Hamiltonian density then becomes $\H_{\k} = \frac{1}{2} \pi_{\k}^2 +
\frac{1}{2} |\k|^2 \phi_{\k}^2$ along with Poisson bracket $\{\phi_{\k},
\pi_{\k'}\} =  \delta_{\k,\k'}$. This is the usual Hamiltonian for decoupled
harmonic oscillator.

In Fock quantization, each of these modes is quantized using Schrodinger method.
Corresponding creation and annihilation operators, namely
$\hat{a}_{\k}^{\dagger}$ and $\hat{a}_{\k}$, are used to express the field
operator and to define \emph{number density} operator $\hat{N}_{\k} =
\hat{a}_{\k}^{\dagger} \hat{a}_{\k}$. Hamiltonian density operator becomes
$\hat{\H}_{\k} = (\hat{N}_{\k}+ \frac{1}{2})|\k|$. In polymer quantization
although the operator $\hat{\H}_{\k}$ exists, the notions of creation and
annihilation operators are not available. Nevertheless, we note that there would
be no Unruh radiation if the acceleration parameter $a$ vanishes. So using
Hamiltonian density operator we can define an alternate number density operator
for Unruh particles of positive frequency $\omegar = \kr >0$, as seen by Rindler
observer, as follows
\begin{equation}\label{Unruh-Number-Operator}
 \hat{\bar{N}}_{\omegar} \equiv  \left[ \hat{\bar{\H}}_{\kr} - 
 \lim_{a\to 0} \hat{\bar{\H}}_{\kr} \right] ~|{\kr}|^{-1} ~.
\end{equation}
This definition of number density operator reproduces the standard result for
Fock quantization.

For the inertial observer we denote the vacuum state as $|0^{\vacl}\rangle
\equiv \prod_{\k} \otimes |0^{\vacl}_{\k}\rangle$ where $\vacl$ refers to a
particular quantization method namely Fock or polymer. Using the equation
(\ref{Hamiltonian:Relation2}) along with the fact that 
$\langle 0^{\vacl}|\hat{\phi}_{\k}|0^{\vacl}\rangle = 0$ and 
$\langle 0^{\vacl}|\hat{\pi}_{\k}|0^{\vacl}\rangle = 0$, 
we express vacuum expectation value of the operator
(\ref{Unruh-Number-Operator}) as
\begin{equation}\label{NumberOperatorGeneralVEV}
 \bar{N}_{\omegar} \equiv 
 \langle 0^{\vacl}| \hat{\bar{N}}_{\omegar} | 0^{\vacl} \rangle
 = \frac{1 - \gamma_{\star}}{e^{2\pi\omegar/a} - 1}  ~,
\end{equation}
where $E_{\k}^0 = \langle 0^{\vacl}|\hat{\H}_{\k}| 0^{\vacl} \rangle$ and 
$\gamma_{\star} = \lim_{\delta\to 0} \gamma_{\star}^{\delta}$ with
\begin{equation}
\label{GammaStarDeltaDef} 
1 - \gamma_{\star}^{\delta} \equiv \frac{1}{\zeta(1+2\delta)}
\sum_{r=1}^{\infty}  \frac{\epsilon_{r}}{r^{1+2\delta}} ~,~~
\epsilon_r \equiv \frac{2 E^0_{\k_r}}{|\k_r|} ~.
\end{equation}

\subsection{Fock quantization}

In Fock quantization, energy spectrum of the $\k^{th}$ oscillator is given by 
$\hat{\H}_{\k} |n_{\k}\rangle = E^n_{\k} |n_{\k}\rangle$ 
with $E^n_{\k} = (n+\frac{1}{2})|\k|$. In this case $\epsilon_r = 1$ which then
implies $\gamma_{\star}^{\delta} = 0$. So the expectation value of the number
density operator in Fock vacuum state $|0^{F}\rangle$ becomes
\begin{equation}\label{NumberOperatorFockVEV}
 \bar{N}_{\omegar} 
=  \langle 0^{F}| \hat{\bar{N}}_{\omegar} | 0^{F} \rangle
= \frac{1}{e^{2\pi\omegar/a} - 1}  ~.
\end{equation}
The equation (\ref{NumberOperatorFockVEV}) represents \emph{thermal} spectrum
for \emph{bosons} at temperature $(a/2\pi k_B)$ where $k_B$ is the
\emph{Boltzmann constant}. This phenomenon is referred as Unruh effect.

\begin{figure}
\includegraphics[width=8cm]{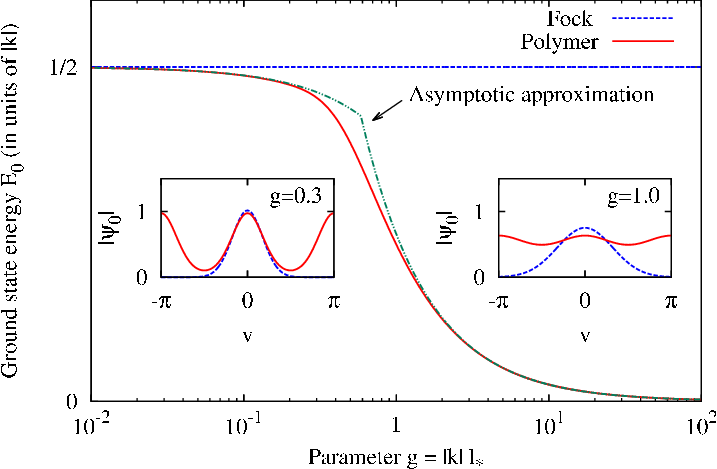}
\caption{\label{fig1}
The comparison of the ground state energy for different modes in Fock and
polymer quantization.  In the left inset panel, the modulus of the polymer
vacuum $ce_0$ mostly follows that of the Fock vacuum, the \emph{Gaussian}, for
smaller $|v|$ ($|\pi_{\k}| << \l^{-1}$). On the right panel, the modulus of
$ce_0$ differs significantly from that of the Gaussian even for smaller $|v|$.
The same relative normalizations for the states have been used in both the
plots.
}
\end{figure}

\subsection{Polymer quantization}

In polymer quantization, one represents exponentiated momentum $U_\lambda \equiv
e^{i\lambda \pi_{\k}}$ as an \emph{elementary} operator in the kinematical
Hilbert space. The classical variables $\phi_{\k}$, $U_\lambda$ satisfy Poisson
bracket $\{\phi_{\k}, U_\lambda\} = i \lambda U_\lambda$. Corresponding
commutator bracket is given by $[\hat{\phi}_{\k},\hat{U}_{\lambda}] = -
\lambda\hat{U}_{\lambda}$.
Operator $\hat{U}_{\lambda}$ is not \emph{weakly continuous} in $\lambda$ hence
$\hat{\pi}_{\k}$ itself cannot be made a well-defined operator. Nevertheless,
one can define an operator $\hat{\pi}_{\k}^{\star} = (\hat{U}_{\l} -
\hat{U}^{\dagger}_{\l})/2i\l$ which can be used to represent the momentum. In
the limit $\l \rightarrow 0$, classically $\pi_{\k}^{\star}$ reduces to
$\pi_{\k}$. In polymer quantization this limit however doesn't exist and $\l$ is
taken to be a \emph{small but finite} scale. This scale is treated as a new
dimension-full parameter of the formulation. In the context of quantum gravity,
one would consider $\l$ to be associated with Planck length $L_{p}$ as $\l \sim
\sqrt{L_{p}}$.

Energy eigenvalues for the $\k^{th}$ oscillator can be computed in polymer
quantization as
\cite{Hossain:2010eb}
\begin{equation}
 \label{EigenValueMCFRelation}
 \frac{E_{\k}^{2n}}{|\k|} = \frac{1}{4g} + \frac{g}{2} ~A_n(g)  ~,~
 \frac{E_{\k}^{2n+1}}{|\k|} = \frac{1}{4g} + \frac{g}{2} B_{n+1}(g) ~,
\end{equation}
where $n\ge0$, $A_n$ and $B_n$ are \emph{Mathieu characteristic value
functions}. Energy eigenfunctions are $\psi_{2n}(v) = \ce_n(1/4g^2,v - \pi/2)$
and $\psi_{2n+1}(v) = \se_{n+1}(1/4g^2,v - \pi/2)$, where $v = \l \pi_{\k} $.
The functions $\ce_n$ and $\se_n$ which are solutions to \emph{Mathieu
equation}, are referred as elliptic cosine and sine functions
\cite{Abramowitz1964handbook}.
The \emph{dimensionless} parameter $g$ is defined as $g = |\k| \l^2 \equiv
|\k|~l_{\star}$ and it signifies the strength of polymer corrections for a given
mode. For low-energy modes \emph{i.e.} for small $g$, energy eigenvalues
(\ref{EigenValueMCFRelation}) become
\begin{equation}
 \label{EEvalueSmallg}
  \frac{E_{\k}^{2n}}{|\k|} \approx \frac{E_{\k}^{2n+1}}{|\k|} \approx
   n+\frac{1}{2} - \frac{(2n+1)^2 + 1}{16} g + \mathcal{O}(g^{2})~.
\end{equation}
Clearly, polymer quantization reproduces known results for low-energy modes. For
trans-Planckian modes \emph{i.e.} for large $g$, ground state energy becomes 
$E_{\k}^0 = \left(1/4g + \mathcal{O}(g^{-3})\right)|\k|$ which is different 
from the expression of low-energy modes.

For simplicity, we extend the asymptotic forms of $E_{\k}^0$ for small and large
$g$, towards $g=(2-\sqrt{2})$ from both ends so that one asymptotic form takes
over from the other asymptotic form continuously in $g$. In other words, we
consider 
$\epsilon_r = (1 - r/\nu_1 r_{\star})$ for $r < r_{\star}$ and 
$\epsilon_r = (r_{\star}/\nu_2 r)$ for $r\ge r_{\star}$ where 
$r_{\star} = (L/l_{\star})(2-\sqrt{2})/2\pi$, $\nu_1 = 2 (2+\sqrt{2})$ and
$\nu_2 = 2 (2-\sqrt{2})$. It can be seen from Fig. (\ref{fig1}) that it is 
a good approximation of exact $\epsilon_r$ and leads to
\begin{equation}
\label{GammaStarPolymer2}
\sum_{r=1}^{\infty}  \frac{\epsilon_{r}}{r^{1+2\delta}} =
\zeta_{r_{\star}}(1+2\delta) 
- \frac{ \zeta_{r_{\star}}(2\delta)}{\nu_1 r_{\star}}
+ \frac{r_{\star} \zeta(2+2\delta, r_{\star})}{\nu_2} ~,
\end{equation}
where $\zeta_{r_{\star}}(1+2\delta) = \sum_{r=1}^{r_{\star}} r^{-(1+2\delta)}$,
 $\zeta_{r_{\star}}(2\delta) = \sum_{r=1}^{r_{\star}} r^{-(2\delta)}$ are 
\emph{truncated zeta functions} and 
$\zeta(2+2\delta, r_{\star}) = \sum_{r=r_{\star}}^{\infty} r^{-(2+2\delta)}$ is 
\emph{Hurwitz zeta function}. In the limit $\delta \to 0$, these functions are
finite as long as $r_{\star}$ is finite. We have mentioned earlier that the
limit $\delta \to 0$ forces the limit $L^{-} \to 0$. However $L^{+}$ and the
polymer scale $l_{\star}$ both being finite, $r_{\star} \sim (L/l_{\star})$
remains finite in the limit $\delta \to 0$. Use of \emph{zeta function identity}
$\lim_{s\to0}[s~\zeta(1+s)]=1$ then leads to $\gamma_{\star} = 1$.

In other words, when one removes the integral regulator $\delta$, 
expectation value of the number density operator (\ref{Unruh-Number-Operator}) 
in polymer vacuum state $| 0^{P} \rangle$ vanishes \emph{i.e.}
\begin{equation}\label{NumberOperatorPolymerVEV}
 \bar{N}_{\omegar} = \langle 0^{P}| \hat{\bar{N}}_{\omegar} | 0^{P} \rangle = 0 
~.
\end{equation} 
% %
% We note here that Fock quantization corresponds to $l_{\star} \to 0$. In 
% this limit \emph{rhs} of the equation (\ref{GammaStarPolymer2}) becomes 
% \emph{regular zeta function} $\zeta(1+2\delta)$ instead of being 
% \emph{truncated}. Together with equation (\ref{GammaStarDeltaDef}), it would 
% then lead to the usual expression (\ref{NumberOperatorFockVEV}). 
% %
We note here that had one taken $l_{\star} \to 0$ (\emph{i.e.} Fock limit) but
keeping the regulator $\delta$ \emph{non-zero}, one would have recovered the 
usual expression (\ref{NumberOperatorFockVEV}). 
% %
Clearly, the limits $\delta\to 0$ and $l_{\star}\to 0$ (or $r_{\star}\to 
\infty$) \emph{do not} commute. 
As mentioned after equation (\ref{Volume-delta-restriction}) that 
$\delta\to 0$ demands $L^{-}\to 0$. This implies that $\delta$ indirectly 
plays the role of an infra-red regulator for Rindler observer given its 
volume is $\Vr = \ln\left(L^{+}/L^{-}\right)/a$. So for Rindler observer, 
removal of \emph{infra-red} regulator leads an infinite sum 
(\ref{GammaStarDeltaDef}) to converge at two different points depending on 
whether \emph{ultra-violet} regulator $l_{\star}$ is present or not. This 
aspect is surprisingly similar to the so-called `UV/IR mixing' phenomena which 
has been extensively studied in \emph{string theory, non-commutative field 
theory} approaches \cite{Minwalla:1999px}.
We note that $r_{\star}\to \infty$ may also be achieved by taking
$L^{+} \to \infty$ although the non-commutative aspect still persists.
Clearly, further studies on these issues are warranted.

In reference \cite{Rovelli:2014gva}, the author has criticized the result 
of this paper. The criticism is based on the claim that since LQG would 
reproduce Fock space two-point function at \emph{zeroth order} then it should 
satisfy KMS condition. Therefore, LQG would predict Unruh effect 
\cite{Rovelli:2014gva}. This argument is \emph{flawed} as ignores all possible 
polymer corrections involving the new scale $l_{\star}$. We have subsequently 
shown \cite{Hossain:2015xqa} that the KMS condition is in fact violated 
in polymer quantization and it is precisely due to the correction terms 
involving the new scale $l_{\star}$. In other word, even following the argument 
of the KMS condition one would conclude that the polymer correction 
causes the particle spectrum to be \emph{non-thermal} which is consistent with 
the disappearance of Unruh effect as shown here.
Furthermore, by computing the response function of an Unruh-DeWitt detector
moving along an uniformly accelerated trajectory, it is shown that induced 
transition rate of the detector contains only transient terms in polymer 
quantization \cite{Hossain:2016klt}. This is in contrast to the response 
function of a similar detector in Fock space where induced transition 
rate contains also a non-transient term and it is proportional to Planck 
distribution at Unruh temperature. Thus, the result shown in 
\cite{Hossain:2016klt} provides an alternative evidence for the result 
presented here.

\section{Discussions}

We have shown that Unruh effect disappears in polymer quantization of
($1+1$) dimensional massless scalar field when one removes the associated 
integral regulator. This result is in contrast to Fock quantization. 
Furthermore, it demonstrates that trans-Planckian modifications can change a 
theoretical prediction. 
In the same note, nevertheless, we would like to point out that there have been
multiple studies in the literature where it is shown that the results from 
polymer quantization of matter fields recover the standard results as obtained 
from Fock quantization in the appropriate limit 
\cite{Hossain:2010eb,Seahra:2012un}. 
In particular, it was shown in \cite{Hossain:2010eb} that polymer quantization 
leads to correct low-energy propagator for scalar field in Minkowski spacetime 
along with perturbative corrections. The polymer corrections are significant 
only for super-Planckian modes. In reference \cite{Seahra:2012un}, it was shown 
that polymer quantization of scalar matter field in Friedmann-Robertson-Walker 
(FRW) background leads to correct scale-invariant power spectra for scalar modes 
with rather small polymer corrections.

We may also point out here that these trans-Planckian modifications are known to 
violate Lorentz invariance \cite{Hossain:2010eb}. So the result shown here may 
not be surprising given one does not expect large deviation from Lorentz 
invariant modifications \cite{Agullo:2008qb}. Therefore, the results as shown 
here indicates a possible deeper relation of Unruh effect with the Planck-scale 
modifications compared to what is often emphasized in the literature and it 
certainly warrants further exploration on the issue. 
Secondly, unlike Fock vacuum state $|0^{F}\rangle$, polymer vacuum state 
$|0^{P}\rangle$ which is represented by \emph{zeroth order elliptic cosine} 
function $\ce_0$ for each mode, effectively mimics the vacuum state even for 
an accelerating observer in the sense that expectation value of the number 
density operator in it remains zero.  
We note that the violation of the Lorentz invariance could lead 
to particle production in a frame moving with a constant velocity in polymer 
quantization as argued in \cite{Husain:2015tna,Kajuri:2015oza}. However, as it 
follows from the equation (\ref{Unruh-Number-Operator}), here we have defined 
the number density operator such that one obtains particle production only due 
to a constant acceleration.

We emphasize that the result shown here is independent of the approximation that
we have made for $\epsilon_{r}$. Rather, it depends on the fact that
$\sum_{r=1}^{\infty} \epsilon_{r} {r^{-(1+2\delta)}}$ is \emph{finite} in
polymer quantization even in the limit $\delta \to 0$.  This is made possible
thanks to the existence of a \emph{small but finite} scale $l_{\star}$ in the
theory. In the context of quantum gravity the scale  $l_{\star}$ would
correspond to Planck length $L_{p}$.
Finally, we note that several proposals have been made in literature to test
Unruh effect in laboratory
\cite{Schutzhold:2006gj,Schutzhold:2008zza,Aspachs:2010hh}. Based on the results
shown here, we may conclude that if experimental measurement of Unruh effect is
ever possible then it can be used to either verify or rule out a theory of
quantum gravity which affects quantization of matter sector due to the presence
of a new scale as described here.

\begin{acknowledgments}
We would like to thank Ritesh Singh for discussions.
GS would like to thank UGC for supporting this work through a doctoral
fellowship.
\end{acknowledgments}

\end{document}